\documentclass[twocolumn,prb,amsmath,amssymb,amsfonts,superscriptaddress,floatfix,aps,showpacs,notitlepage]{revtex4}
\usepackage{graphicx}
\usepackage{float}
\usepackage{graphicx}
\usepackage{longtable}
\usepackage{dcolumn}
\usepackage{bm}
\usepackage{color}
\usepackage{tabularx}

\usepackage{dcolumn}
\usepackage{bm}

\newcommand{\be}{\begin{equation}}
\newcommand{\ee}{\end{equation}}
\newcommand{\bea}{\begin{eqnarray}}
\newcommand{\eea}{\end{eqnarray}}

\begin{document}
\title{\emph{Ab initio} calculation of the effective Coulomb interactions in $MX_2$ ($M$=Ti, V, Cr, Mn, Fe, Co, Ni; $X$=S, Se, Te): intrinsic magnetic ordering and Mott insulating phase}

\author{A. Karbalaee}
\affiliation{Department of Physics, Faculty of Science, University of Zanjan, Zanjan 45371-38791, Iran}
\author{S. Belbasi}
\affiliation{Department of Physics, Faculty of Science, University of Zanjan, Zanjan 45371-38791, Iran}
\author{H. Hadipour}
\email{hanifhadipour@gmail.com}
\affiliation{Department of Physics, University of Guilan, 41335-1914, Rasht, Iran}

\date{....}


\begin{abstract}
Correlated phenomena such as magnetism and Mott phase are a very controversial issue in two-dimensional transition
 metal dichalcogenides (TMDCs). With the aim of finding the value of correlation strength and understanding the origin of
 ferromagnetic order in TMDCs, we first identify relevant low-energy degrees of freedom on both octahedral T and trigonal prismatic H lattices in $MX_2$ ($M$=Ti, V, Cr, Mn, Fe, Co, Ni; $X$=S, Se, Te) and then determine the strength of the effective Coulomb interactions between localized $d$ electrons from the first principles using the constrained random-phase approximation.
 The on-site Coulomb interaction (Hubbard \emph{U}) values lie in the range 1.4$-$3.7 eV (1.1$-$3.6 eV) and depend on
the ground-state electronic structure, \emph{d}-electron number, and correlated subspace.
For most of the TMDCs we obtain $1 < U/W_b < 2$ (the bandwidth $W_b$), which turn out to be larger than the corresponding values in elementary
transition metals. On the basis of the calculated $U$ and exchange $J$ interaction, we have checked the condition to be fulfilled for the formation of the ferromagnetic order by Stoner criterion.
The results indicate that experimentally observed Mn$X_2$ ($X$=S, Se) and
V$X_2$ ($X$=S, Se) have an intrinsic ferromagnetic behavior
in pristine form, although V-based materials
are close vicinity to the critical point separating ferromagnetic
from paramagnetic phase.
\end{abstract}

\pacs{.....}

\maketitle

\section{Introduction}\label{sec1}

Electron correlation has been an important issue in low-dimensional systems after the experimental synthesis of graphene \cite{Geim,Katsnelson} and other two-dimensional (2D) materials \cite{Huang,Gong,yu}.
Generally 2D materials have less bandwidth than the bulk due to lower continuity of the density of states, as a consequence, strength of the correlation $U/W_{b}$, namely the ratio of effective electron-electron interaction $U$ to bandwidth $W_{b}$, becomes great \cite{Sasoglu}.
  For 2D compounds containing transition metal (TM) atoms, in addition to quantum confinement effects arising from reduced dimensionality, correlation effects are expected to play a crucial role in determining the electronic and magnetic properties due to the presence of narrow $t_{2g}$ or $e_g$ states at the vicinity of Fermi level \cite{Yekta}. One of the consequences of a moderate correlation $U/W_{b}\sim 1$ or strong correlation $U/W_{b}\gg 1$ are inducing magnetic ordering \cite{Yasin,Yekta}, Mott insulator\cite{Yekta,Isaacs,Wang-3}, and etc in low-dimensional systems. For instance, an intrinsic ferromagnetism in the monolayers of TM halides such as Cr$X_3$ ($X$=Cl, Br, I)\cite{McGuire,Zhang,Siena,Seyler,Shcherbakov,Jiang}, VI$_3$ \cite{Son,Kong},  NiI$_2$\cite{Kurumaji} has been observed experimentally and confirmed theoretically \cite{Besbes,Tian,He,Liu,Torelli}. Furthermore, due to the large correlation strength $U/W_b>2$, most of the TM halides exhibit a Mott insulating character especially in the systems with nearly half-filled 3$d$ bands \cite{Yekta}. Besides TM halides, other 2D materials such as in Cr$_2$Ge$_2$Te$_6$ \cite{Gong}, Fe$_3$GeTe$_2$ \cite{Deng,Fei},  V$X_2$ ($X$=Se,Te)\cite{Li,Duvjir,Bonilla,Guo,Ma,Wang-2}, TiTe$_{2}$\cite{Lasek}, MnSe$_2$ \cite{OHara}, Cr$X_2$ ($X$=S, Se, Te)\cite{Houlong,Freitas,Sun-1,Purbawati,Habib,Li-x,Meng-1},  Co$X_2$($X$=S, Se, Te) have been found to show magnetic orderings.

In the mentioned systems, synthesis of the 3$d$ TM dichalcogenides (TMDCs) such as $MX_2$ ($M$=V, Cr, Mn; $X$=S, Se, Te) has led to a huge experimental and theoretical interest due to the existence of incredibly rich correlated phenomena ranging from room temperature intrinsic ferromagnetism \cite{yu,Lasek,Bonilla,Guo,Ma,OHara,Wang-2} to charge density wave ordering\cite{Chen-1,Sugawara,Feng-1}, and Mott phase \cite{Isaacs}. Experimentally, even at room temperature, ferromagnetic order were reported in Mn and V-based $MX_2$ compounds\cite{Bonilla,OHara}. The possibility of 100$\%$ spin polarization indicates that $MX_2$ monolayers are promising material for spintronics devices.
 Also, including the on-site Coulomb interaction leads to Mott insulating behavior for V-based systems due to the presence of a half-filled band\cite{Isaacs}.
  However, for most of  the 3$d$ TMDCs such as Cr$X_2$, Mn$X_2$ systems and even V$X_2$, the origin of the intrinsic ferromagnetism remained controversial and existence of strongly correlated phase is still in doubt \cite{Fumega,Wong,Feng-1,Wong-1,Duvjir,Chen-1,Vinai}.
 In other words, some other opposite studies indicate that the ferromagnetism
 seen in the TMDCs is extrinsic 2D magnetism stemming from vacancies\cite{Chua-1}, or proximity effects \cite{Zhang-2,Vinai} which are not able to be precisely eliminated in 2D crystal growth.
 The formation of such extrinsic magnetic moments and long-range magnetic order induced by atom vacancies in some other 2D nonmagnetic materials such as graphene \cite{Yazyev,Hadipour,Sasioglu-1,Ugeda} and MoS$_2$ \cite{Zhang-1,Cai} makes absence of intrinsic 2D magnetism for pristine TMDCs more likely. From the theoretical side, magnetic ordering depends strongly on the correlated subspace of the $d$-shell and its value of
 electron correlation \cite{Wong}. Anyway, controversy exists over the intrinsic magnetism in the monolayer of 3$d$ TMDCs.
Another interesting property of 3$d$ TMDCs such as trigonal prismatic phase of Cr$X_2$ and V$X_2$ is valley polarization\cite{Houlong,Tong}.
Most of the valleytronic $MX_2$ have a appropriate spin-polarized
bandgap of 1.0 eV and spontaneous valley polarization are found to be about 40$-$90 meV \cite{Tong,Chengan},
large enough for valleytronic devices.
 The presence of both large valley polarization and spin splitting in a
single material are attractive for the research on valleytronic and spintronic applications.

To prevail over the problems of band gap underestimation in Density functional theory (DFT) based on the local density approximation (LDA) or generalized gradient approximation (GGA) and to improve band dispersion to find proper valley polarization,  GWA (in particular self-consistent GW + spin orbit coupling) was performed to induce many-body correction \cite{Fuh,Thygesen,Filip,Chen-1,Cazzaniga}.
However, due to presence of narrow $d$ bands in TMDCs monolayer,  the refined methods such as DFT+$U$ and DFT plus dynamical mean-field theory (DFT + DMFT) will be needed. DFT +$U$ method has been employed to study the electronic properties of some TMDCs \cite{Esters,Houlong-1,Wang-1}, in which the effective $U$ parameters are usually taken from the values found for other materials including the same transition metal atom.
Only a few works have been performed the $ab initio$ linear response theory and constraint random phase approximation (cRPA)\cite{Cococcioni,Aryasetiawan,Nomura} to calculate Hubbard $U$  for VS$_2$\cite{Isaacs,SchoNhoff-1}. Indeed, the results of this approach are not extensive and usually are not consistent with those extracted from experiments. This motivates us to do fully \emph{ab initio} calculation to find effective Coulomb parameters for all 3$d$ TMDCs monolayers.  The $U$  parameters obtained from $ab initio$ calculations not only provide a fundamental understanding of the correlated phenomena in TMDCs,  but these effective interaction can also be used in model Hamiltonians thus increasing the predictive power of model calculations.

In this systematic study we first identify appropriate correlated subspaces by constructing Wannier function and then by employing the cRPA approach within the full-potential linearized augmented-plane-wave (FLAPW) method, we determine the effective on-site Coulomb interaction of the $d$ electrons in $MX_2$ ($M$=Ti, V, Cr, Mn, Fe, Co, Ni; $X$=S, Se, Te) for both H and T phases.
We find $1 < U/W_b < 2$ in most 3$d-MX_2$, making them moderately correlated
materials. In non-magnetic state, among the metallic TMDCs, the correlation strength $U/W_b$ turn out to be  2.7
in T-Mn$X_2$ and 1.8 in H-V$X_2$ with values much larger than the corresponding values in elementary TMs.
Based on $U$ and exchange interaction $J$ values, we discuss the tendency of the
electron spins to order ferromagnetically by Stoner criterion.
The results indicate that experimentally observed Mn$X_2$ ($X$=S, Se) and
V$X_2$ ($X$=S, Se) have an intrinsic ferromagnetic behavior
in pristine form. Moreover, V-based materials are close
to the edge of paramagnetic to ferromagnetic transition

\begin{figure}[b]
	\centering
	\includegraphics[width=90mm]{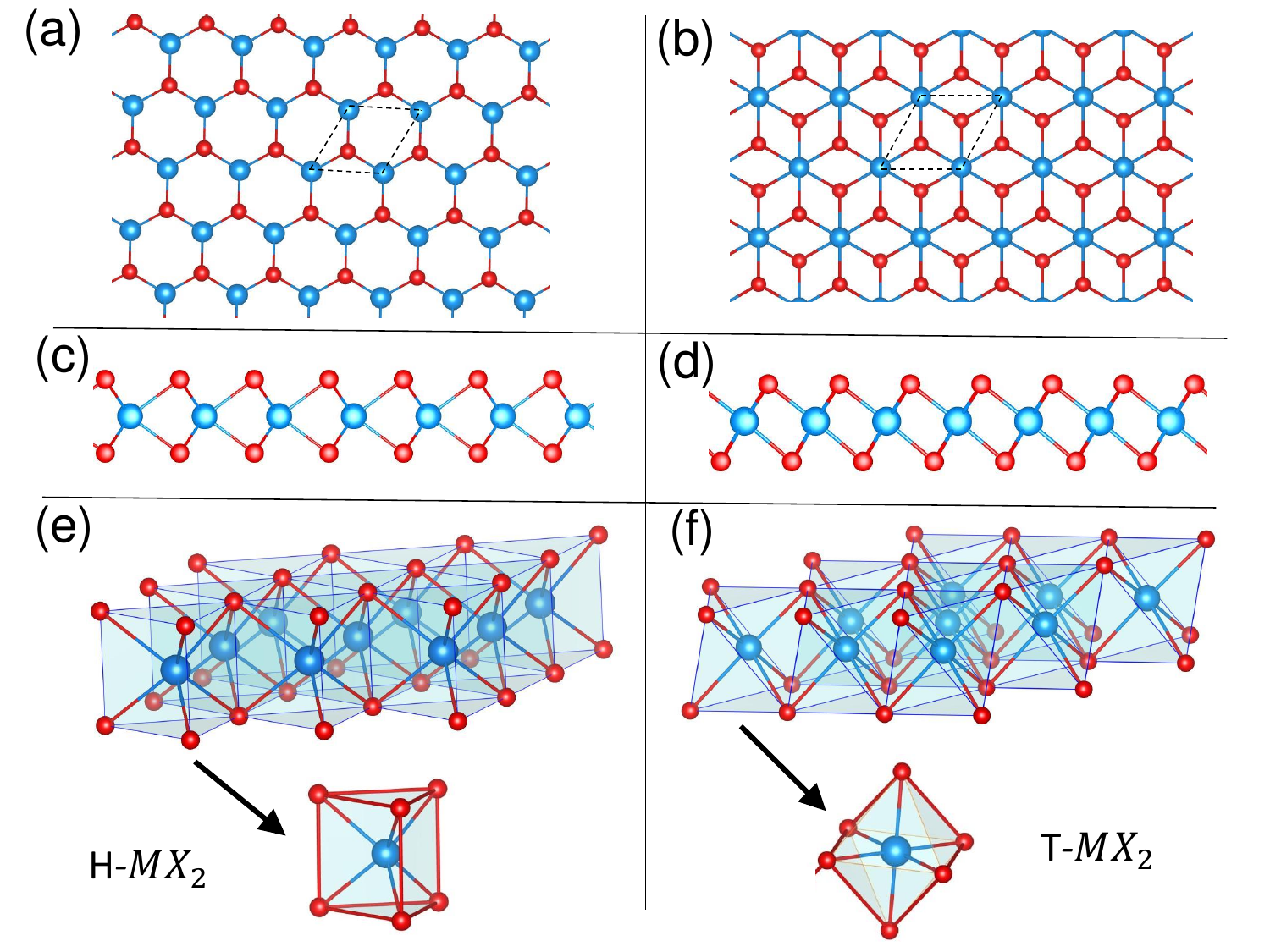}	
	\vspace{-0.1 cm}
	\caption{Top view of the two-dimensional crystal of $MX_2$ for (a) H phase (b) T phase. Side view of $MX_2$ for (c) H phase (d) T phase.
     (e) Trigonal prismatic coordination of one TM and six chalcogen \emph{X} atoms in H-phase.
     (f) Octahedral coordination in T phase where in each TM atoms are bound to six \emph{X} atoms. The blue and red spheres exhibit $M$ and $X$ atoms, respectively}
	\label{fig:subm1}
\end{figure}

\section{Computational details}\label{sec2}
\begin{figure}[t]
	\centering
	\includegraphics[width=80mm, height=100mm]{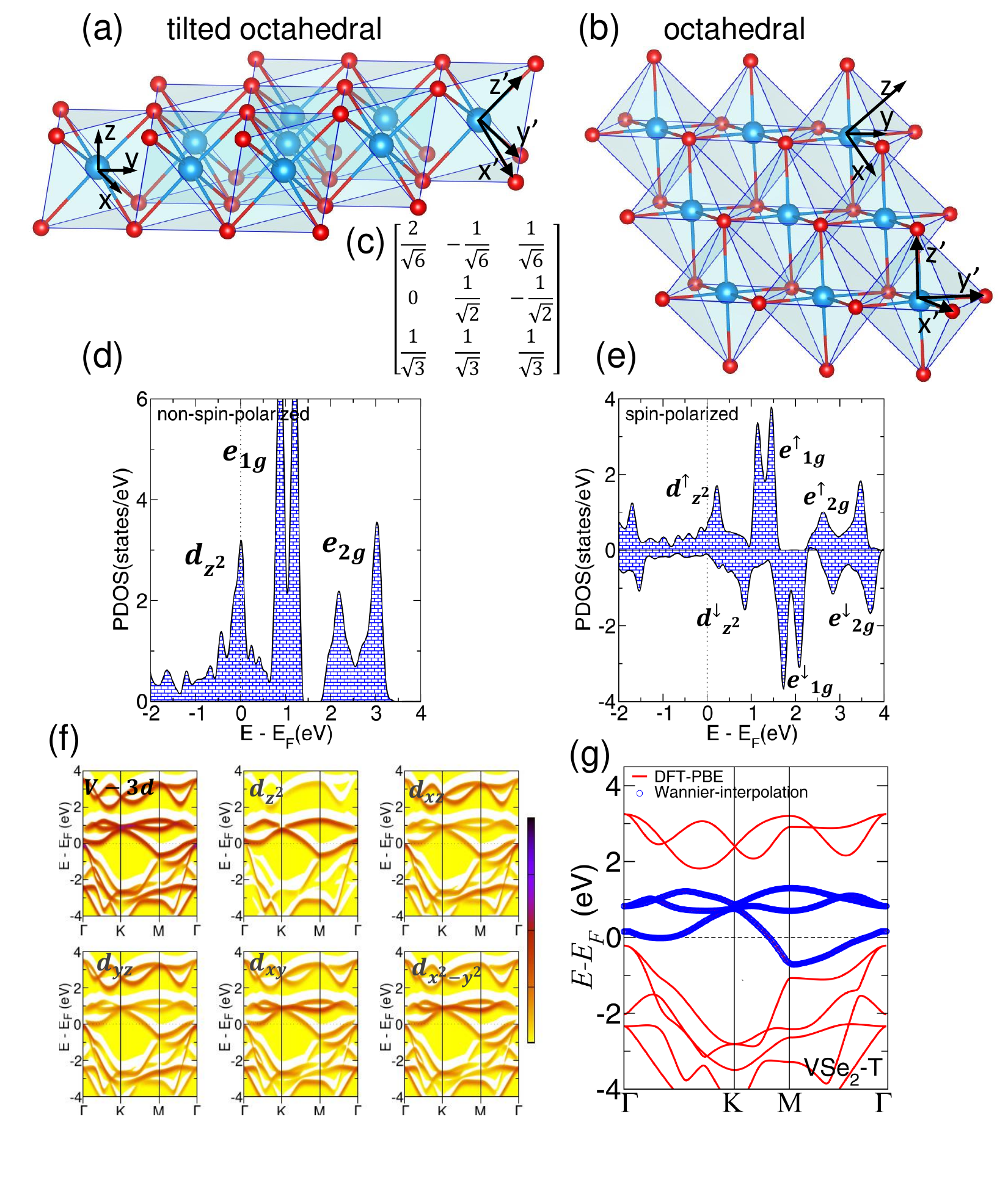}
	\vspace{-0.4 cm}
	\caption{(a) Tilted octahedral crystal structure  with respect
to the standard $x, y, z$ Cartesian coordinate in T phase where in two opposite faces of the octahedron are parallel to
the layers. (b) The same structure in the rotated state $x', y', z'$ so that it forms a conventional octahedral phase. (c) The rotation matrix
  which convert  $x, y, z$ to $x', y', z'$ coordinate.
(d) DOS projected onto 3$d$ states of the $V$ atom in the non-spin-polarized calculation for T-$VSe_2$.
(e) The same as (d) for spin-polarized calculation. (f) The orbital-projected
band structures for 3$d$ electron of V atom of T-$VSe_2$ based on DFT-PBE. (g) DFT-PBE (red) and Wannier interpolated band structures (blue) of T-$VSe_2$ monolayer using $d_{z^{2}}+e_{g1}$ subspace.}
	\label{fig:subm2}
\end{figure}

There are two common types of crystal structures in monolayer of TMDCs, which are trigonal prismatic (H) and octahedral (T) coordination \cite{Li-1}. The side and top views of the crystal structures
of T and H phase of $MX_2$ systems are presented in Fig.~\ref{fig:subm1}(a-d). The difference in crystal field splitting generated by surrounding chalcogen $X$ atoms for two lattice of H-$MX_2$ and T-$MX_2$ leads to different correlated subspace and play a key role in expressing the differences in the observed electronic and magnetic properties.
So, we first briefly describe the crystal structural of $MX_2$ in T and H phases.
The lattice of T-$MX_2$ consists of triangular nets of TM atoms so that the $X$ atoms are arranged as octahedra
with $M$ atoms in the center (see Fig.~\ref{fig:subm1}(f)) . In this arrangement, $d$ electrons states splits into three lower-energy orbitals $t_{2g}$ ($d_{xy}$, $d_{xz}$, $d_{yz}$) and two higher-energy orbitals $e_g$ ($d_{z^{2}}$, $d_{x^{2}-y^{2}}$). Note that the octahedron is tilted with respect to the standard Cartesian coordinate $x, y, z$
system, in such a way that $C_4$ axes (the axes goes through two opposite vertices of the octahedron) is in the $z'$ direction and
two opposite faces of the octahedron are parallel to the layers. As shown in Fig.~\ref{fig:subm2}(a) and ~\ref{fig:subm2}(b), in fact,  one of the eight triangles of octahedron is lying on the floor and the $z$ axis is perpendicular to this triangle. In this situation, the linear combination of the local $t'_{2g}$ orbitals, namely $d_{x'y'}$, $d_{x'z'}$, $d_{y'z'}$
will be the low energy levels. Using the rotation matrix reported in Fig.~\ref{fig:subm2}(c), the singlet $d_{z^{2}}$ orbital which is oriented
perpendicular to the layer, is given by $(d_{x'y'}+d_{x'z'}+d_{y'z'})/\sqrt{3}$. The other $t'_{2g}$ orbitals, $(d_{y'z'}-d_{x'y'})/\sqrt{2}$, and $(d_{y'z'}+d_{x'y'}-2d_{x'z'})/\sqrt{6}$, are corresponding to $e_{1g}$ doublet states.  The two  $d_{z'^{2}}/d_{x'^{2}-y'^{2}}$  are denoted by high-energy degenerate doublets $e_{2g}$ states.
By these notation, the non-spin polarized density of $d$ state for T-$VSe_2$ are presented in Fig.~\ref{fig:subm2}(d).
Note that the bands are not of pure $d_{z^{2}}$, $e_{1g}$, and $e_{2g}$ character but are mixtures. Thus, this nominations like $d_{z^{2}}$ refer to their dominant orbital character.

\begin{figure}[t]
	\centering
	\includegraphics[width=80mm, height=100mm]{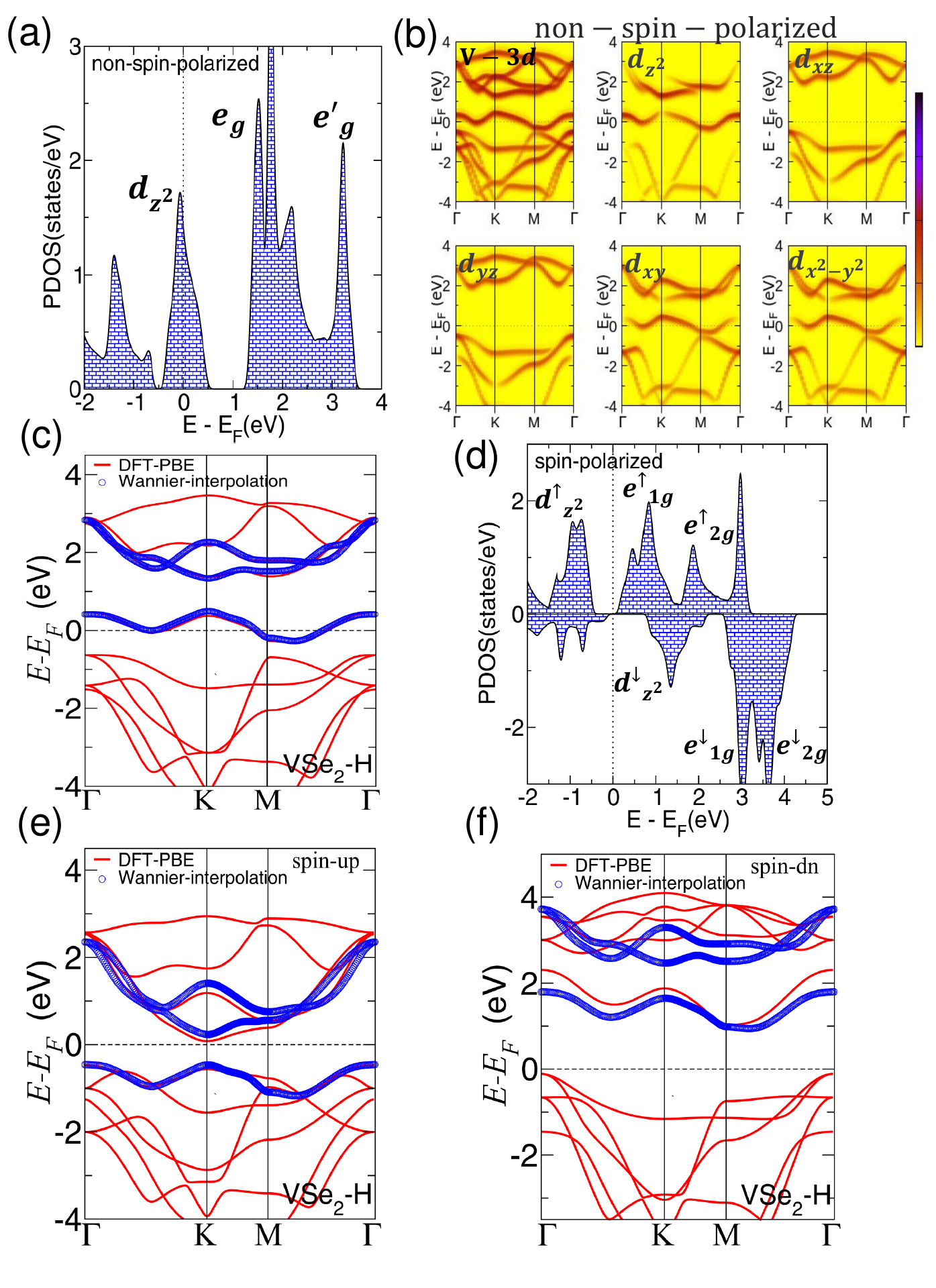}
	\vspace{-0.4 cm}
	\caption{(a) DOS projected onto 3$d$ states of the $V$ atom in the non-spin-polarized calculation for H-$VSe_2$.
            (b) The orbital-projected band structures for 3$d$ electron of V atom of H-$VSe_2$ based on DFT-PBE.
            (c) DFT-PBE (red) and Wannier interpolated band structures (blue) of H-$VSe_2$ monolayer using $d_{z^{2}}+e_{g}$ subspace.
            (d) The same as (a) for spin-polarized calculation.  DFT-PBE and Wannier interpolated band structures of H-$VSe_2$ monolayer using $d_{z^{2}}+e_{g1}$ subspace for (e) spin up and (f) spin down.}
	\label{fig:subm3}
\end{figure}

For H-$MX_2$, because of the trigonal prismatic coordination Fig.~\ref{fig:subm1}(e),
 in the crystal-field level, $d$-shell splits into a singlet $d_{z^{2}}$, a low-energy doublet $e$ ($d_{xy}$ and $d_{x^{2}-y^{2}}$) and a high-energy doublet $e'$ ($d_{xz}$, $d_{yz}$).
According to this notation, in Fig.~\ref{fig:subm3}(a) we have presented the non-spin polarized density of $d$ state for H-phase of H-$VSe_2$.

The basic unit cell is hexagonal in all TMDCs and consists of three atoms.
We consider $MX_2$ ($M$=Ti, V, Cr, Mn, Fe, Co, Ni; $X$=S, Se, Te) for both T
and H phases. $MX_2$ unit cells containing one $M$ and two $X$ atoms, are simulated based on the
slab model having a 25 $\AA$ vacuum separating them.
For the non-spin-polarized
as well as for the spin-polarized DFT calculations, the FLAPW method as implemented in the FLEUR
code \cite{Marzari,Freimuth,Fleur} is used.

We employ GGA in the Perdew-Burke-Ernzerhof (PBE) parameterization \cite{Perdew} for
the exchange-correlation energy functional.
Since, Coulomb matrix elements are almost sensitive to the internal coordinates
of the atoms and structural distortions, for all calculations the lattice parameters and atomic positions
are chosen to be equal to the optimized parameters with
considering the relaxation of the atomic coordinates and possible distortions.
The $z$ coordinate of two $X$ atoms are optimized with the residual force less than
0.01 eV/\AA. In the scf calculation, $16\times16\times1$ $k$-point grids are used for unit cells of all systems.
A linear momentum cutoff of $G_{max}$= 4.5 bohr$^{-1}$
is chosen for the plane waves.
The maximally localized
Wannier functions (MLWFs) \cite{Marzari} are constructed with the
Wannier90 library using appropriate bands per $M$ atom.
The DFT calculations
are used as an input for the SPEX code \cite{Mostofi,Friedrich} to determine the strength of
 Coulomb interaction between correlated
electrons from the cRPA and RPA methods \cite{Aryasetiawan,Nomura}.
A dense $k$-point grid $12\times12\times1$ are used in the cRPA and RPA calculations.

To find strength of screened Coulomb interaction
we need to identify correlated subspace which helps to construct Wannier
functions properly. So, at first we calculate
both non-magnetic and magnetic orbital-resolved density
of states (DOS) for all systems which will discussed in
next section by details. Here, we have only presented the
orbital projected band structure of VSe$_2$ in two phases
,namely T-VSe$_2$ and H-VSe$_2$, in Fig.~\ref{fig:subm2}(f) and Fig.~\ref{fig:subm3}(b) respectively
that resolve the contribution of different $d$ states.
The reason  why we have chosen VSe$_2$ case is that the origin of the ferromagnetism and Mott phase
in it's monolayer limit is still under debate.
 Also,  it was one of the first material in TMDCs,
 in which ferromagnetism is detected experimentally in monolayer.

As shown in Fig.~\ref{fig:subm2}(f) in T phase, in the non-magnetic calculation the mixture of $d_{z^{2}}$ and $e_{1g}$ ($d_{xy}/d_{x^{2}-y^{2}}$) is significant which is also observed in total DOS of $d$ states (see Fig.~\ref{fig:subm2}(d)) but $t'_{2g}$ ($d_{z^{2}}+e_{1g}$) states  are  well-isolated  bands at the vicinity of Fermi level. Although,
we construct Wannier functions individually for three  $d_{z^{2}}$, $t'{_2g}$, and the full $d$ shell orbitals
as a correlated subspace. In $t'_{2g}$ and $d$ cases, the original and
the Wannier-interpolated bands agree very well. In Fig~\ref{fig:subm2}(g), we have presented a comparison of the non-spin-polarized DFT-PBE
band structures with the corresponding Wannier-interpolated
bands obtained with the $t'_{2g}$ Wannier orbitals as a minimal subspace
for T phase of VSe$_{2}$. In other words, it motivates a three-band model for VSe$_2$ with $t'_{2g}$ Wannier orbitals.
In a same way, correlated subspaces are defined for all considered T phase of TMDCs which will be discussed in the next section.

In the H-VSe$_2$, a single half-filled band which is predominantly $d_{z^{2}}$ orbital character
is well separated from the other bands in the non-magnetic calculation.
Thus, to find minimal correlated subspace in the H phase we construct Wannier functions
for the $d_{z^{2}}$, $e_{g}$ orbitals.
These energy levels are mainly responsible
for the electronic, magnetic, and transport properties of H-VSe$_2$.
Note that $e'_{g}$ states are far from Fermi level and do not contribute to construct half-filled band at E$_F$.
The comparison of the DFT-PBE band structure with Wannier interpolation (see Fig.~\ref{fig:subm3}(c))indicates that the best
consistency can be obtained by $d_{z^{2}}+e_g$ Wannier orbitals as a minimal subspace.

Note that, since one of the aim of this paper is to determine effective Coulomb parameters for
low-energy model Hamiltonian of correlated TM materials, the results before symmetry breaking take place, such as
non-magnetic $U$ should be calculated. Despite this, we have done spin-polarized calculation for a few systems.
As shown in Fig.~\ref{fig:subm2}(f) and ~\ref{fig:subm3}(a), the crystal field splitting is almost small in TMDCs materials. So, considering spin polarization may cause to have problems with entangled bands. It complicates the construction of Wannier function if we use $d_{z^{2}}+e_g$ subspace in the magnetic calculation. As shown in the Fig.~\ref{fig:subm3}(e) and ~\ref{fig:subm3}(f), the original and
the Wannier-interpolated bands do not agree very well.
In this case we must go beyond the $d_{z^{2}}+e_g$ minimal subspace, for instance, full $d$-shell subspace, which wrongly eliminate a few specific states in the screening.

In the following, we briefly describe cRPA method.
The fully screened Coulomb interaction $\tilde{U}$
is related to the bare Coulomb interaction $V$ by
\begin{equation}
\tilde{U}(\boldsymbol{r},\boldsymbol{r}',\omega)=\int d\boldsymbol{r}''
\epsilon^{-1}(\boldsymbol{r},\boldsymbol{r}'',\omega) V(\boldsymbol{r}'',\boldsymbol{r}'),
\label{fullysw}
\end{equation}
where $\epsilon(\boldsymbol{r},\boldsymbol{r}'',\omega)$
is the dielectric function. The dielectric function  is related to
the electron polarizability $P$ by
\begin{equation}
\epsilon(\boldsymbol{r},\boldsymbol{r}',\omega)=\delta(\boldsymbol{r}-\boldsymbol{r}')-\int d\boldsymbol{r}'' V(\boldsymbol{r},\boldsymbol{r}'')P(\boldsymbol{r}'',\boldsymbol{r}',\omega),
\label{rpadiel1}
\end{equation}
where the RPA polarization function $P(\boldsymbol{r}'',\boldsymbol{r}',\omega)$ is given by
\begin{equation}
\begin{gathered}
P(\boldsymbol{r},\boldsymbol{r}',\omega)= \\
2 \sum_{m}^\mathrm{occ} \sum_{m'}^\mathrm{unocc} \varphi_{m}(\boldsymbol{r}) \varphi_{m'}^{*}(\boldsymbol{r}) \varphi_{m}^{*}(\boldsymbol{r}') \varphi_{m'}(\boldsymbol{r}')  \\
\times\Bigg[ \frac{1}{\omega-\Delta_{mm'}+i\eta} - \frac{1}{\omega+\Delta_{mm'}-i\eta} \Bigg].
\end{gathered}
\label{rpapol1}
\end{equation}
Here, $\varphi_{m}(\boldsymbol{r})$ are the single-particle DFT Kohn-Sham eigenfunctions,
and $\eta$ a positive infinitesimal. $\Delta_{mm'}=\epsilon_{m'}-\epsilon_{m}$ with the Kohn-Sham eigenvalues
$\epsilon_{m}$. ß

In the cRPA approach, in order to exclude the screening due to the correlated subspace, we
separate the full polarization function of Eq.~(\ref{rpapol1}) into two parts
\begin{equation}
P=P_{d}+P_{r},
\label{rpapol2}
\end{equation}
where $P_{d}$ includes only the transitions ($m\rightarrow m'$) between the states of the correlated subspace
and $P_{r}$ is the remainder. Then, the frequency-dependent effective Coulomb interaction
is given schematically by the matrix equation
\begin{equation}
U(\omega) = [1-VP_{r}(\omega)]^{-1}V.
\end{equation}
It contains, in $P_r$, screening processes that would not be
captured by the correlated subspace and excludes the ones that
take place within the subspace.

The matrix elements of the effective Coulomb interaction in the MLWF basis are
given by
\begin{equation}
\begin{gathered}
U_{\mathbf{R}n_{1},n_{3},n_{2},n_{4}}(\omega)= \\
\int\int d\boldsymbol{r}d\boldsymbol{r}' w_{n_{1}\mathbf{R}}^{*}(\boldsymbol{r}) w_{n_{3}\mathbf{R}}(\boldsymbol{r}) U(\boldsymbol{r},\boldsymbol{r}',\omega) w_{n_{4}\mathbf{R}}^{*}(\boldsymbol{r}') w_{n_{2}\mathbf{R}}(\boldsymbol{r}'),
\end{gathered}
\label{hubudef31}
\end{equation}
where $w_{n\mathbf{R}}(\boldsymbol{r})$ is the MLWF at site $\mathbf{R}$ with orbital index
$n$, and the effective Coulomb potential $U(\boldsymbol{r},\boldsymbol{r}',\omega)$ is calculated
within the cRPA as described above. We define the average Coulomb matrix elements  $U$, $U'$,
and $J$ in the static limit ($\omega=0$) as follows \cite{Anisimov1,Anisimov2}:
\begin{equation}
U=\frac{1}{L}\sum_{m}
U_{mm;mm}\,,
\label{U_diag}
\end{equation}
\begin{equation}
U'=\frac{1}{L(L-1)}\sum_{m \neq n}
U_{mn;mn}\,,
\label{U_offdiag}
\end{equation}
\begin{equation}
J=\frac{1}{L(L-1)}\sum_{m \neq n}U_{mn;nm}\,,
\label{Hund_J}
\end{equation}
where $L$ is the number of localized orbitals, i.e., two for $e_{g}$ and three for $t_{2g}$ orbitals.
This parametrization of partially screened Coulomb interactions is the so-called Hubbard-Kanamori
parametrization. Similar to the definition of $U$ ($U'$, $J$), we can also define the so-called
fully screened interaction parameters $\tilde{U}$  ($\tilde{U'}$, $\tilde{J}$)  as well
as unscreened (bare) $V$. The bare $V$ provides information about the localization of Wannier functions
and is a useful parameter in the interpretation of the screened Coulomb interaction parameters.

\begin{figure}[t]
	\centering
	\includegraphics[width=90mm, height=70mm]{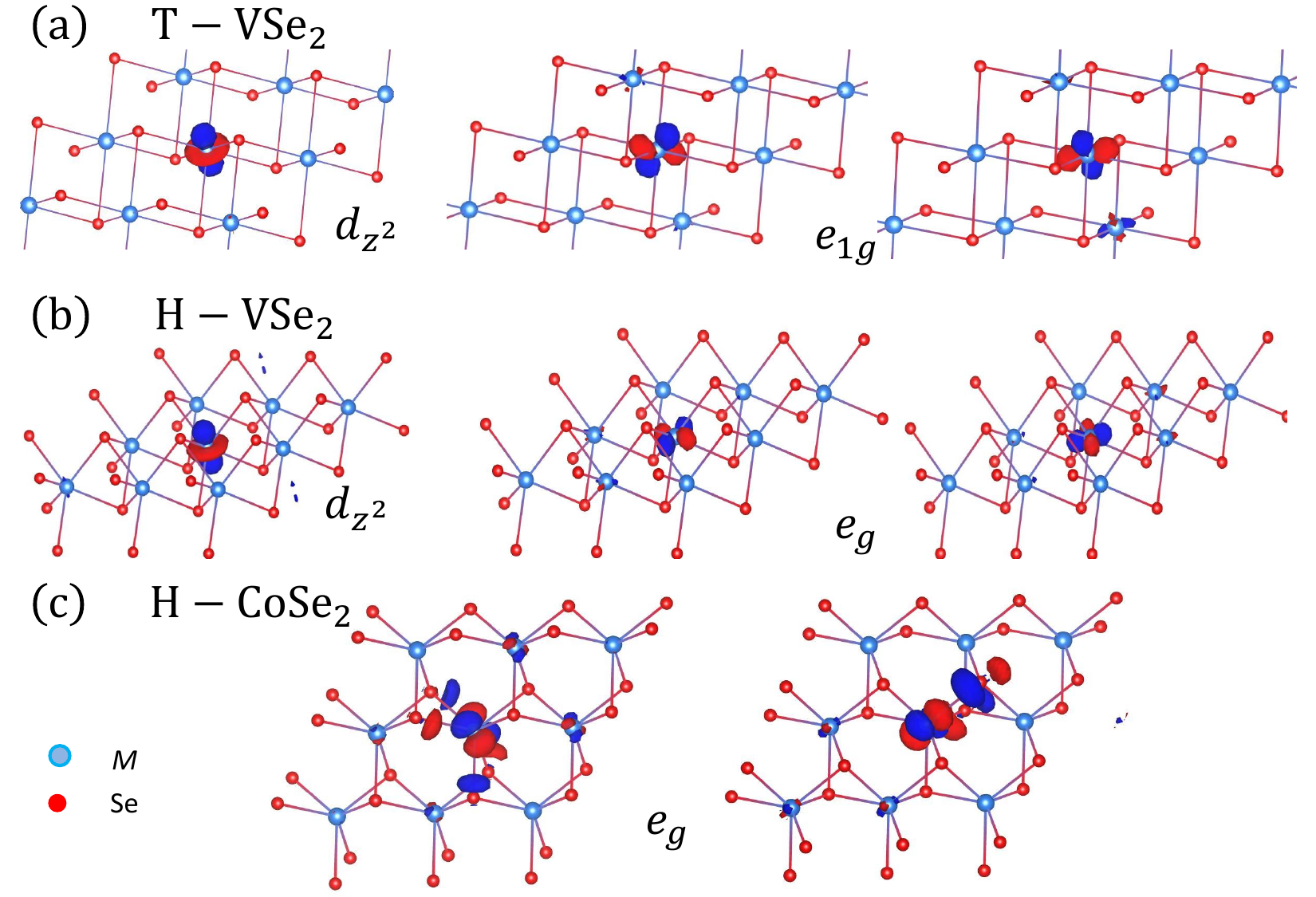}
	\vspace{-0.4 cm}
	\caption{(a) The $d_{z^{2}}$ and $e_{1g}$-like MLWF for V atom of T-VSe$_2$. (b) The $d_{z^{2}}$ and $e_{g}$-like MLWF for V atom of H-VSe$_2$. (c) The $e_{g}$-like MLWF for Co atom of H-CoSe$_2$}
	\label{fig:Wannier}
\end{figure}

\section{Results and discussion}\label{sec3}

\begin{figure*}[t]
	\centering
	\includegraphics[width=180mm, height=120mm]{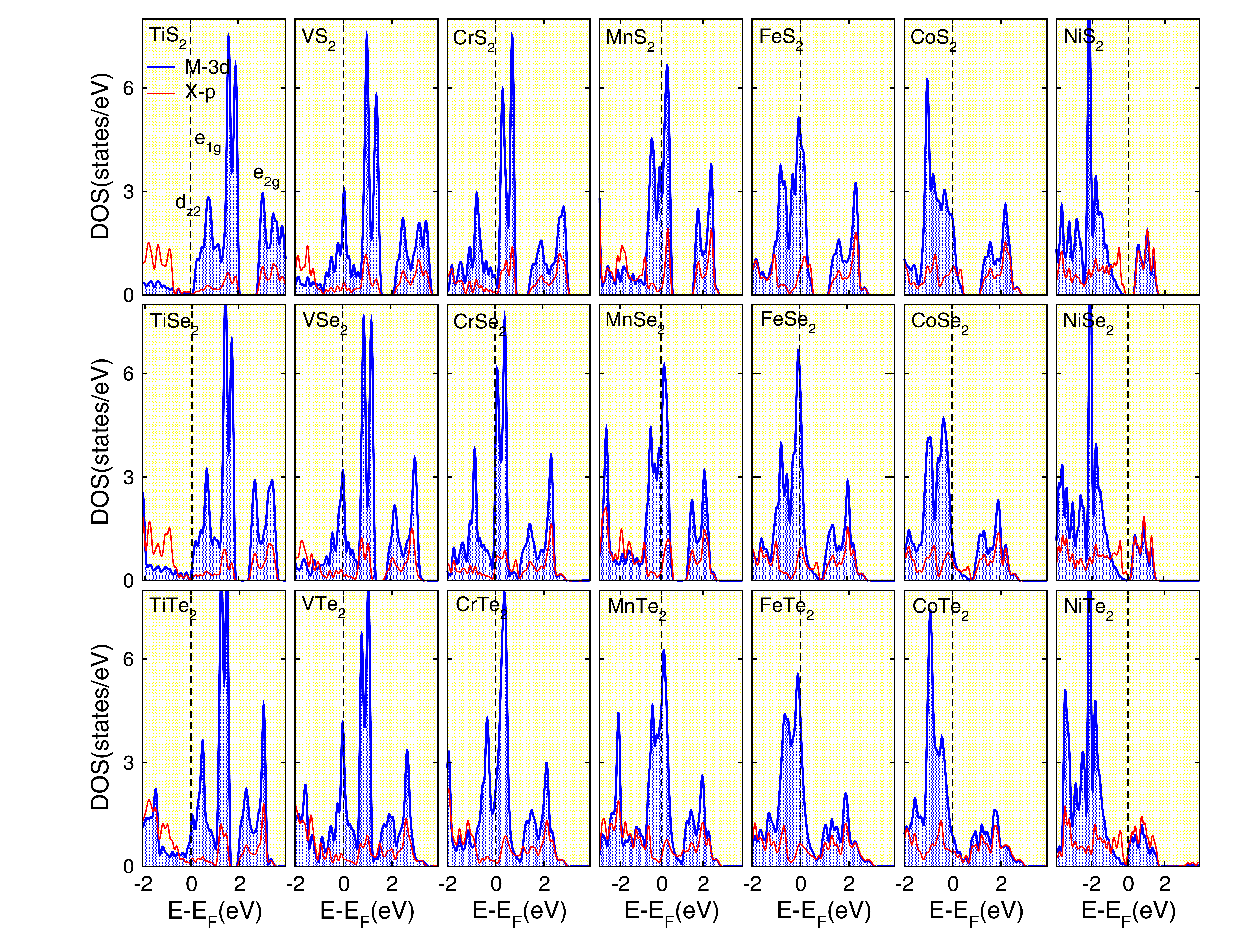}
	\vspace{-0.4 cm}
	\caption{DOS projected onto 3$d$ states of the $M$ atom as well as on $p$ states of the $X$ atoms for T-$MX_2$ materials.
            The three distinct peaks of $d$ bands correspond to $d_{z^{2}}$, $e_{1g}$ and $e_{2g}$ states.}
	\label{fig:subm4}
\end{figure*}

We start with the discussion of appropriate correlated subspace moving
from early Ti$X_2$ to late Ni$X_2$ for both T and H phases.
To define correlated subspace and also to identify the contribution of different $M$ atoms in
the screening, in Fig.~\ref{fig:subm4} we present non-magnetic orbital-resolved DOSs for all considered M$X_2$ monolayers in T-phase.
Similar to V-based system, the comparison of the DFT-PBE band structure with Wannier
interpolation shows that for all compounds except Ni-based T-M$X_2$, the bands with $t'_{2g}$ ($d_{z^{2}}$+$e_{1g}$) character is the well-defined correlated subspace. This results is to be expected because of the existence of large $t'_{2g}$ states near the E$_F$ for Ti$X_2$ to Co$X_2$.
In the case of T-Ni$X_2$ systems (see Fig.~\ref{fig:subm4}(h) for example), the Fermi level are located in the energy gap between $t'_{2g}$ and $e_{2g}$ bands and the minimal correlated subspace depends on the type of electron or hole doping. The original and the Wannier-interpolated bands do not agree very well when we consider  subspace $t'_{2g}$. For these systems, we have defined $e_{2g}$,
and full $d$ states as correlated subspace. It means $e_{2g}$ is well-defined subspace in zero-doping, but for
optical properties and other correlated phenomenon $d$ subspace would be necessary.
As shown in Fig.~\ref{fig:subm4}, the orbital-resolved DOS of T-$MX_2$ with $X$= S, Se, Te look very similar,
thus we have determined Hubbard $U$ parameters identically moving from $X$=S to Te systems.
The Fig.~\ref{fig:subm4} exhibits a strong admixture of chalcogen $p$ with $t'_{2g}$ states which increase as one moves from
$M$=Ti to Ni-based systems and also from $X$=S to Te ones.
To find correlated subspace in these cases, we included a few more states in construction of the Wannier functions.
We find $U$ matrix elements of for example $p$-admixed $t'_{2g}$ states and pure $t'_{2g}$
are nearly identical. However, the delocalization effect arising from admixture of $p$ states will be reflected in the value of coulomb parameters later.

\begin{figure*}[t]
	\centering
	\includegraphics[width=180mm, height=50mm]{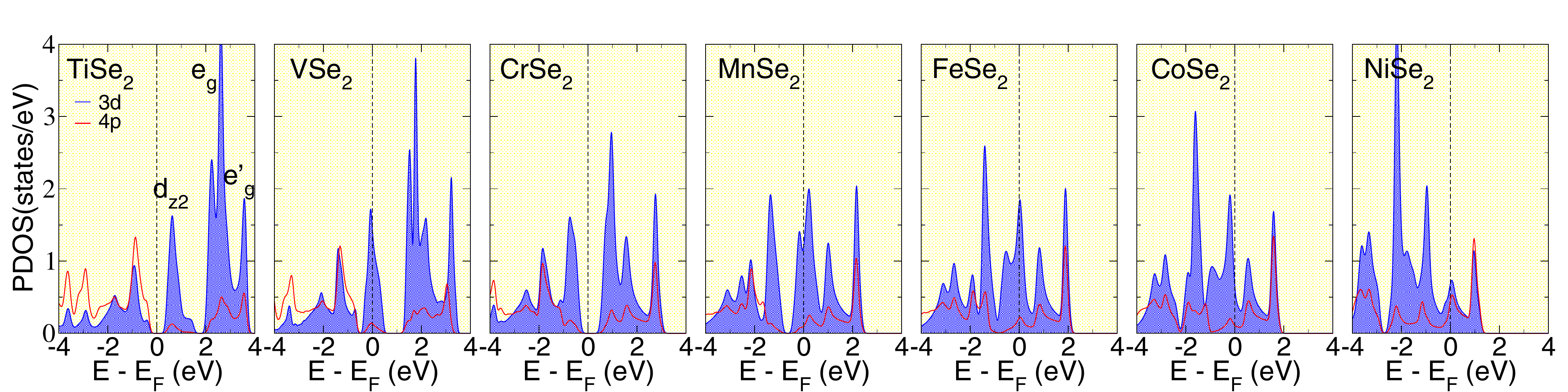}
	\vspace{-0.4 cm}
	\caption{DOS projected onto 3$d$ states of the $M$ atom as well as on $p$ states of the $Se$ atoms for H-$MSe_2$ materials.
            The three distinct peaks of $d$ bands correspond to $d_{z^{2}}$, $e_{g}$ and $e'_{g}$ states.}
	\label{fig }
	\label{fig:subm5}
\end{figure*}

For the H phase of $MX_2$, the situation for defining the proper subspace is more complicated and differ from one system to another system.
The behavior of the electronic structure of 3$d$ orbitals across the TM atoms, from $X$=S to Te,
is similar to the case of T-phase. So, we only depicted the orbital resolved DOS only for $M$Se$_2$ materials in Fig.~\ref{fig:subm5}.
For Ti- and V-based $MX_2$, although single band with predominantly $d_{z^{2}}$ character
is well separated from the other states, but the best consistency are given by $d_{z^{2}}+e_g$ states.
However, because of the great importance of single $d_{z^{2}}$ band which almost allows defining an effective single band
low-energy Hamiltonian, we have also reported the electron-electron interactions for this single band correlated subspace.
In $M$=Cr- to Fe-based $MX_2$ cases, we see the substantially contribution of both $d_{z^{2}}$ and $e_g$ states
to the DOS around the Fermi energy and, as a consequence, bands with predominantly $d_{z^{2}}+e_g$ character are obtained
to construct Wannier function properly.
For the last two series, namely H-Co$X_2$ and H-Ni$X_2$, we find $e_g$ and $e'_g$ respectively as the minimal correlated subspaces and  these layers can be described by a two-orbital effective low-energy model. Although a four-orbital
$e_g+e'_g$ low-energy model can be used for the Ni case with caution.

Now, we come to the part we discuss the value of average on-site bare (unscreened) Coulomb interaction
$V$, the average on-site partially (fully) screened
interaction parameters $U$ ($\tilde{U}$), as well as exchange interaction $J$($\tilde{J}$) for correlated subspace electrons of $MX_2$ systems.
The results for T-$MX_2$ and H-$MX_2$ are presented in Table ~\ref{table:1} and ~\ref{table:2}  respectively.
In both H and T phases, bare interaction $V$ increases with increasing the number of electrons in the $d$-shell.
This is to be expected because moving
from the left Ti to the right Ni in periodic table, the nuclear charge
increases which leads to the contraction of $d$-wave functions, and subsequently increase the localization of Wannier functions.
Note that, the values of V reach a maximum for Fe$X_2$ systems and drop off sharply
for Co$X_2$. For a better understanding, we plot the shape of Wannier orbitals for early VSe$_2$ and late CoSe$_2$ systems in Fig. ~\ref{fig:Wannier}.
As shown in Fig. ~\ref{fig:Wannier}(c), the coupling of e$_g$ states to neighboring chalcogene $p$ states is
significant, which leads to a delocalization and, therefore, to smaller
$V$ interaction. Furthermore, the results for the
chalcogene series $MX_2$ with $X$=S to Te shows a reduction in bare $V$, which is well understood by the increase
in the lattice constant, making the Wannier function more extended.

The calculated $U$ values for T-$MX_2$ (H-$MX_2$) lie between 1.4 and 3.7 eV (1.1 and 3.6 eV) and depend on the ground-state
electronic structure, $d$-electron number, chalcogene $X$, and correlated subspace.
Since, all screening channels with an excitation energy larger than a few eV around E$_F$
are contributed in cRPA calculations of $MX_2$ systems, we see strong screening with large difference $V-U$ values in
these materials.  As seen in Table ~\ref{table:1} and ~\ref{table:2}, the $U$ for $M$ sites tend to increase when $M$ is varied from Ti to Fe. In each chalcogen $X$ and a particular subspace, Hubbard $U$ parameter is determined by two effects: i) Wannier localization effect due to increasing $d$-electron number and ii) electronic structure effects. Similar to the bare interaction, the first effect
is important also in Hubbard $U$ which tend to enhance from Ti to Ni.
For the second effect, the insight from orbital-resolved DOS of T-phase (H-phase) depicted in Fig.~\ref{fig:subm4} (Fig.~\ref{fig:subm5}) is that
below the $d$ states  there is a broad peak of chalcogen $p$ orbitals,
which should contribute with $p \rightarrow t'_{2g}$ ($p \rightarrow d_{z^{2}}+e_g$) transitions substantially to the screening.
Indeed, as we move from Ti to Ni, the chalcogen $p$ states gradually move to the lower energy
and accompanied by TM 3$d$ bands. So, the contribution of $p \rightarrow d$ transition into the polarization
function does not change and even slightly reduces when across the series Ti to Fe and give rise to the enhancement of
of $U$ parameters with increasing 3$d$ electron number in both phase of TMDCs.
Furthermore, there is another important screening processes via $t'_{2g} \rightarrow e_{2g}$ ($d_{z^{2}}+e_g \rightarrow e'_g$) in T-phase (H-phase) of $MX_2$ systems,
which only reduce the values of Coulomb parameters and does not change the trend across the TM series.

SchoNhoff et al. \cite{SchoNhoff-1} determined Hubbard $U$ values for $V$S$_2$ ($V$Se$_2$) by employing the cRPA method.
The obtained $U$ values are 2.25 eV (2.40 eV), which are smaller than
the calculated Coulomb interactions with $d_{z^{2}}+e_g$ subspace 3.12 eV (2.96 eV) and larger than the Coulomb interactions
 with $d_{z^{2}}$ subspace 1.51 eV (1.43 eV) presented in Table ~\ref{table:2}.
Using the linear response approach, Isaacs et al. calculated  $U$ values
  for H phase of VS$_2$  in non-magnetic state and obtained $U$ = 4.14 eV \cite{Isaacs}.
  This value is almost 1.0 eV larger than Hubbard interaction in Table ~\ref{table:2}.
  For the magnetic state, they also determined $U$ interactions for both T and H phases
  and obtained 3.84 eV for H phase and 3.99 eV for T phase.
 What subspace to use in construction of the Wannier basis or how to define as the Hubbard $U$ parameter
 are the possible reasons for this disagreement.

\begin{table*}[!ht]
	\caption{Lattice parameter, orbital type of correlated subspace, bandwidth $W_b$, on-site intraorbital bare $V$, intraorbital partially (fully) screened $U(\tilde{U})$, interorbital partially (fully) screened $U'(\tilde{U'})$, partially (fully) screened exchange interaction $J(\tilde{J})$, correlation strength $U/W_b$, and the DOS at the Fermi level $D(E_F)$ for T-$MX_2$ compounds.}
\centering
\begin{ruledtabular}
\begin{tabular}{ccccccccccccc}
 $MX_2$& $a({\AA})$ & Orbitals & $W_b$(eV) & $V$(eV)  & $U(\tilde{U}$)(eV) & $U'(\tilde{U'})$(eV) & $J(\tilde{J})$(eV) & $U/W_b$ & $D(E_F)$ \\ \hline
		TiS$_2$  &  3.4176 & $d_{z^2}$+$e_{1g}$ & 1.85 & 15.12 & 2.22(1.56) & 1.41(0.98) & 0.40(0.29) & 1.20 & 0.07 \\
		TiSe$_2$ &  3.5441 & $d_{z^2}$+$e_{1g}$ & 1.81 & 14.69 & 1.85(1.27) & 1.04(0.67) & 0.40(0.29) & 1.02 & 0.38 \\
		TiTe$_2$ &  3.7409 & $d_{z^2}$+$e_{1g}$ & 1.98 & 13.08 & 1.36(0.96) & 0.68(0.49) & 0.33(0.22) & 0.69 & 1.15 \\
		VS$_2$	 &  3.1917 & $d_{z^2}$+$e_{1g}$ & 2.17 & 16.25 & 2.56(0.93) & 1.69(0.48) & 0.43(0.22) & 1.18 & 3.12 \\
		VSe$_2$  &  3.3341 & $d_{z^2}$+$e_{1g}$ & 1.92 & 16.09 & 2.14(0.83) & 1.29(0.40) & 0.42(0.21) & 1.11 & 3.26 \\
		VTe$_2$  &  3.6020 & $d_{z^2}$+$e_{1g}$ & 2.34 & 14.47 & 1.77(0.63) & 1.02(0.29) & 0.36(0.16) & 0.76 & 4.31 \\
		CrS$_2$  &  3.0656 & $d_{z^2}$+$e_{1g}$ & 2.07 & 16.68 & 2.78(1.12) & 1.93(0.16) & 0.42(0.21) & 1.35 & 0.30 \\
		CrSe$_2$ &  3.2243 & $d_{z^2}$+$e_{1g}$ & 2.16 & 16.14 & 2.49(0.60) & 1.75(0.12) & 0.38(0.17) & 1.16 & 5.80 \\
		CrTe$_2$ &  3.6932 & $d_{z^2}$+$e_{1g}$ & 2.12 & 15.59 & 2.27(0.48) & 1.52(0.30) & 0.37(0.14) & 1.07 & 4.23 \\
		MnS$_2$  &  3.3507 & $d_{z^2}$+$e_{1g}$ & 1.12 & 18.03 & 3.02(0.39) & 2.06(0.09) & 0.49(0.15) & 2.70 & 4.81 \\
		MnSe$_2$ &  3.4912 & $d_{z^2}$+$e_{1g}$ & 1.19 & 17.71 & 2.79(0.32) & 1.88(0.06) & 0.45(0.12) & 2.35 & 5.63 \\
		MnTe$_2$ &  3.7448 & $d_{z^2}$+$e_{1g}$ & 2.38 & 15.46 & 2.22(0.30) & 1.45(0.07) & 0.36(0.11) & 0.93 & 5.92 \\
		FeS$_2$  &  3.2013 & $d_{z^2}$+$e_{1g}$ & 1.51 & 18.34 & 3.13(0.60) & 2.19(0.37) & 0.46(0.11) & 2.07 & 4.13 \\
		FeSe$_2$ &  3.3682 & $d_{z^2}$+$e_{1g}$ & 1.69 & 17.98 & 2.81(0.56) & 1.86(0.32) & 0.46(0.12) & 1.66 & 4.59 \\
		FeTe$_2$ &  3.6269 & $d_{z^2}$+$e_{1g}$ & 2.55 & 12.80 & 1.63(0.44) & 1.07(0.26) & 0.29(0.08) & 0.64 & 3.61 \\
		CoS$_2$  &  3.2281 & $d_{z^2}$+$e_{1g}$ & 1.54 & 13.19 & 2.06(1.02) & 1.58(0.50) & 0.42(0.25) & 1.34 & 2.20 \\
		CoSe$_2$ &  3.3704 & $d_{z^2}$+$e_{1g}$ & 2.63 & 12.13 & 1.93(0.74) & 1.50(0.40) & 0.30(0.15) & 0.73 & 1.31 \\
		CoTe$_2$ &  3.6227 & $d_{z^2}$+$e_{1g}$ & 2.72 & 10.58  & 1.57(0.59) & 1.06(0.32) & 0.24(0.12) & 0.58 & 0.98 \\
		NiS$_2$  &  3.3583 & $e_{2g}$           & 1.75 & 18.23 & 3.05(2.30) & 2.06(1.56) & 0.48(0.38) & 1.74 & 0.00 \\
		         &         & $d$                & 3.58 & 21.46 & 3.67(2.62) & 2.41(1.33) & 0.63(0.54) & 1.02 & 0.00 \\
		NiSe$_2$ &  3.5492 & $e_{2g}$           & 1.86 & 18.01 & 2.16(1.80) & 1.02(0.82) & 0.58(0.48) & 1.16 & 0.00 \\
		         &         & $d$                & 3.49 & 20.28 & 3.09(2.39) & 1.97(1.06) & 0.61(0.50) & 0.88 & 0.00 \\
		NiTe$_2$ &  3.7806 & $e_{2g}$           & 2.13 & 17.65 & 1.88(1.42) & 0.97(0.71) & 0.45(0.35) & 0.88 & 0.00 \\
                 &         & $d$                & 4.32 & 18.45 & 2.65(1.87) & 1.62(0.92) & 0.53(0.41) & 0.61 & 0.00 \\
\end{tabular}
\label{table1}
\end{ruledtabular} \label{table:1}
\end{table*}

\begin{table*}[!ht]
	\caption{The same as Table ~\ref{table:1} for H phase of $MX_2$.}
\centering
\begin{ruledtabular}
\begin{tabular}{ccccccccccccc}
 $MX_2$& $a({\AA})$ & Orbitals & $W_b$(eV) & $V$(eV)  & $U(\tilde{U}$)(eV) & $U'(\tilde{U'}$)(eV) & $J(\tilde{J}$)(eV) & $U/W_b$ & $D(E_F)$ \\ \hline
		TiS$_2$ & 3.3376  & $d_{z^2}$+$e_{g}$ & 2.23 &  14.47 & 2.88(2.62) & 2.02(1.74) & 0.52(0.41) & 1.29 & 0.00 \\
                &         & $d_{z^2}$         & 1.20 &  13.92 & 1.83       & ------     & ------     & 1.53 & 0.00 \\
		TiSe$_2$& 3.4739  & $d_{z^2}$+$e_{g}$ & 2.39 &  14.18 & 2.55(2.26) & 1.70(1.32) & 0.51(0.39) & 1.07 & 0.00 \\
                &         & $d_{z^2}$         & 1.05 &  13.26 & 1.62       & ------     & ------     & 1.54 & 0.00 \\
		TiTe$_2$& 3.7277  & $d_{z^2}$+$e_{g}$ & 2.78 &  13.43 & 2.03(1.59) & 1.25(0.88) & 0.47(0.36) & 0.73 & 0.00 \\
                &         & $d_{z^2}$         & 0.96 &  13.09 & 1.27       & ------     & ------     & 1.32 & 0.00 \\
		VS$_2$	& 3.1650  & $d_{z^2}$+$e_{g}$ & 2.97 &  15.05 & 3.12(1.16) & 2.23(0.59) & 0.55(0.34) & 1.05 & 1.52 \\
                &         & $d_{z^2}$         & 0.95 &  14.49 & 1.51       & ------     & ------     & 1.59 & 0.65 \\
		VSe$_2$ & 3.3066  & $d_{z^2}$+$e_{g}$ & 3.02 &  14.73 & 2.94(0.81) & 2.01(0.33) & 0.53(0.31) & 0.97 & 1.80 \\
                &         & $d_{z^2}$         & 0.76 &  14.24 & 1.43       & ------     & ------     & 1.88 & 0.62 \\
		VTe$_2$& 3.5458   & $d_{z^2}$+$e_{g}$ & 3.40 &  14.05 & 2.45(0.65) & 1.40(0.25) & 0.49(0.29) & 0.72 & 1.72 \\
                &         & $d_{z^2}$         & 0.69 &  13.40 & 1.12       & ------     & ------     & 1.62 & 0.51 \\
        CrS$_2$ & 3.0552  & $d_{z^2}$+$e_{g}$ & 3.41 &  15.53 & 3.18(1.96) & 2.28(1.43)	& 0.56(0.34) & 0.93 & 0.00 \\
		CrSe$_2$& 3.2108  & $d_{z^2}$+$e_{g}$ & 3.62 &	15.31 &	2.93(1.67) & 2.01(1.15) & 0.53(0.33) & 0.81 & 0.00 \\
		CrTe$_2$& 3.4319  & $d_{z^2}$+$e_{g}$ & 3.90 &  14.49 & 2.36(1.21) & 1.55(0.88)	& 0.52(0.33) & 0.61 & 0.00 \\        	
		MnS$_2$ & 3.0944  & $d_{z^2}$+$e_{g}$ & 3.09 &	17.11 &	3.46(0.82) & 2.42(0.50) & 0.62(0.26) & 1.12 & 1.78 \\
		MnSe$_2$& 3.2335  & $d_{z^2}$+$e_{g}$ & 3.32 &	16.92 &	3.10(0.68) & 2.08(0.42) & 0.61(0.24) & 0.93 & 1.65 \\
		MnTe$_2$& 3.5164  & $d_{z^2}$+$e_{g}$ & 3.68 &	16.32 &	2.68(0.53) & 1.79(0.32) & 0.58(0.23) & 0.73 & 2.05 \\
        FeS$_2$ & 3.1426  &	$d_{z^2}$+$e_{g}$ & 3.13 &  18.59 &	3.58(1.14) & 2.45(0.67) & 0.67(0.25) & 1.14 & 1.75 \\
		FeSe$_2$& 3.3040  & $d_{z^2}$+$e_{g}$ & 3.18 &  18.46 &	3.31(1.06) & 2.19(0.62) & 0.67(0.24) & 1.04 & 1.94 \\
		FeTe$_2$& 3.5238  & $d_{z^2}$+$e_{g}$ &	3.29 &  17.68 &	2.74(0.95) & 1.76(0.54) & 0.61(0.22) & 0.83 & 1.81\\
        CoS$_2$ & 3.2037  & $e_{g}$	          & 2.45 &	13.40 &	1.88(0.67) & 1.40(0.44) & 0.44(0.29) & 0.77 & 2.07 \\
		CoSe$_2$& 3.3406  & $e_{g}$           &	2.51 &	12.93 &	1.72(0.61) & 1.26(0.38) & 0.43(0.28) & 0.68 & 1.96 \\
		CoTe$_2$& 3.5485  & $e_{g}$           &	2.73 &	11.50 &	1.59(0.53) & 1.11(0.36) & 0.43(0.28) & 0.58 & 2.18 \\
	    NiS$_2$ & 3.4782  & $e_{g}$+$e'_{g}$  & 2.96 &	16.99 &	3.07(0.99) & 2.36(0.75) & 0.45(0.32) & 1.04 & 0.74 \\
                &         & $e'_{g}$	      & 1.31 &	16.23 &	1.92(0.75) & 1.24(0.62) & 0.45(0.31) & 1.47 & 0.42 \\
		NiSe$_2$& 3.4263  &$e_{g}$+$e'_{g}$   & 3.15 &	15.17 &	2.61(0.82) & 1.94(0.63) & 0.39(0.31) & 0.83 & 0.67\\
                &         & $e'_{g}$	      & 1.43 &	14.38 &	1.66(0.57) & 0.92(0.49) & 0.37(0.27) & 1.16 & 0.39 \\
		NiTe$_2$& 3.6122  &$e_{g}$+$e'_{g}$   &	3.23 &	13.22 &	1.91(0.62) & 1.40(0.47) & 0.36(0.27) & 0.59 & 0.69 \\
                &         & $e'_{g}$	      & 1.60 &	12.81 &	1.17(0.41) & 0.55(0.28) & 0.35(0.26) & 0.73 & 0.37 \\
\end{tabular}
\label{table1}
\end{ruledtabular} \label{table:2}
\end{table*}

Note that exclusive correlated subspaces must be incorporated
into the $U$ calculation, thus, becomes meaningless to
compare the Coulomb matrix elements results of H-$MX_2$ ($M$=Ti,V,Co,Ni) and T-Ni$X_2$ monolayers with the corresponding results of other
$MX_2$ compounds. For instance, considering the single band $d_{z^{2}}$ as a correlated subspace for H-phase of VSe$_2$,
there is an extra sizable screening channels via $d_{z^{2}} \rightarrow e_g$ transition which can reduce
Hubbard $U$ with respect to $U$ for the $d_{z^{2}}+e_g$ subspace.
On the other hand, the value of $U$=1.43 eV for H-VSe$_2$ is smaller than
the $U$=1.62 eV for H-TiSe$_2$, which does not follow the increasing trend across the TM series.
It can be described by the $M-3d$ and $X-p$ projected DOS around the Fermi energy in Fig.~\ref{fig:subm5}.
Indeed, as we go from Ti to V in H-phase of $MX_2$, the  Se-4$p$
states move towards $d_{z^{2}}$ resulting
in more contribution into the polarization
function. It compensates the increase in $U$ caused by Wannier localization, leading to the reduction
of $U$ interaction with increasing 3$d$ electron number. The same behavior is also observed in S- and
Te-based H-$MX_2$ materials (see Table~\ref{table:2}).

On going from S to Te within each $MX_2$ systems,
the lattice constant increase, as a consequence, the longer
bond lengths lead to smaller orbitals overlap.
It can bring the states  closer together energetically as shown in Fig.~\ref{fig:subm4} for T phase, and thus smaller energy difference increase
the contribution of $p \rightarrow t'_{2g}$ and $t'_{2g} \rightarrow e_{2g}$ transitions ($p \rightarrow d_{z^{2}}+e_g$ and $d_{z^{2}}+e_g \rightarrow e'_g$ transition in H-$MX_2$) into the polarization functions.
This means that Coulomb screening is enhanced in the $M$S$_2$-$M$Se$_2$-$M$Te$_2$ sequence and thus
effective Coulomb interaction $U$ reduce in the same sequence in both T and H phase.
The situation for compounds with other
correlated subspaces is almost the same.
For instance, considering the $d_{z^{2}}$ correlated subspace the $U$ value is reduced in VTe$_2$ with respect to the VS$_2$ case.

In Table ~\ref{table:1}, moving again from Ti to Ni, the same increasing trend is observed for partially inter-orbital Coulomb interactions $U'$. The obtained $J$ parameters for $MX_2$ vary in the range 0.24-0.67 eV. Despite the very different range of values in respect to $U$ and $U'$, exchange $J$ gradually increase with the $d$-electron number, which are less affected by the electronic structures.
In cubic symmetry, Coulomb matrix elements, namely $U$ and $U'$, and $J$  fulfills the relation $U'=U-2J$.
Even though TMDCs do not have cubic symmetry, this relation is nearly satisfied in most of them.

So far, we focus on the partially screened
Coulomb interactions $U$.  To analysis the screening within the correlated subspace, we also
calculate fully screened Coulomb interactions $\tilde{U}$ and are
reported within parentheses for T-$MX_2$ (H-$MX_2$) in Table~\ref{table:1} (Table~\ref{table:2}).
Except for the semiconductors T-Ni$X_2$, H-Ti$X_2$, and H-Cr$X_2$, the efficient metallic screening
even causes more than 80$\%$ difference between $U$ and $\tilde{U}$ parameters.
Actually, the calculated $\tilde{U}$ parameters depend strongly on
the density of state at the vicinity of E$_F$.
For example, as seen in Table~\ref{table:1}, the $\tilde{U}$ value for
VSe$_2$ in T-phase (H-phase) is 0.83 eV (0.81 eV)
being about 60$\%$ smaller than $U$=2.14 eV (2.94) due to the large DOS at E$_F$ and subsequently significant
 contribution of the $t'_{2g} \rightarrow t'_{2g}$ ($d_{z^{2}}+e_g \rightarrow d_{z^{2}}+e_g$) transitions to the polarization function.
On the opposite side, in H-Ti$X_2$, T-Ni$X_2$, H-Cr$X_2$, the screening within the correlated subspace is very weak due to
the presence of the band gap in the electronic structures,  and $(U-\tilde{U})/U$ reaches to 15-25$\%$ in these materials.
Therefore, the behavior of fully screened interactions $\tilde{U}$
is completely different from $U$, as they do not follow the
 ordering with respect to increase of $d$-electron number.

In the following, we discuss the strength of the electronic correlations,
namely the ratio of the effective Coulomb interaction $U$ to the bandwidth $W_b$ ($U/W_b$) in TMDCs.
Note that the $U/W_b$ values are determined for a non-spin-polarized
state. Let us start by the results for T phase presented in Table ~\ref{table:1}.
For most of the H-$MX_2$ compounds, we find  $1 < U/W_b < 2$  which put these systems in
 the moderately correlated regime. Therefore, Coulomb interaction play an important
  role in model Hamiltonian study of the correlation effects in T-phase of TMDCs which induce
  correlated phenomena like magnetic order.
Starting from T-Ti$X_2$, the correlation strength
$U/W_b$ increases and reaches to the maximum in Mn-based TMDCs, and then tends to decrease to Co-based systems.
Moreover, electron correlation is strong $U/W_b > 2$  in  T-Mn$X_2$ ($X$=S, Se) materials, as a results,
beside magnetic ordering, it becomes unstable to Mott phase.

The $U/W_b$ for the H-phase of TMDCs does not show any clear trend with significant fluctuation around $U/W_b=1$
from one system to another.  We find that in H-$MX_2$ ($M$=Ti, V) with $d_{z^{2}}$ correlated subspace,
the correlation strength $U/W_b$ is maximum, bringing them closer to the Mott phase.

It is interesting to compare our calculated $U/W_b$ values with reported ones for elementary TMs and other layered
materials containing TM atoms. While, the obtained $1 < U/W_b < 2$ values for TMDC compounds turn out to be larger than
the corresponding values in elementary transition metals $U/W_b < 1$\cite{Sasoglu},
they are significantly smaller than the ones calculated for TM-halides $U/W_b > 2$ \cite{Yekta}.
$U/W_b \sim 1$ were found in the case of $M_2$C and $M_2$CO$_2$ MX-enes, which reveals moderate electronic
correlations in these materials\cite{Yasin}. Note that although the \emph{ab initio} Coulomb parameters of TMDCs are smaller than
elementary 3$d$ TMs, MX-enes, the narrow bands with $d_{z^{2}}$, $e_{1g}$, and $e_{g}$ character presented in TMDCs result
 in a larger $U/W_b$  correlation strength.

\begin{figure}[t]
	\centering
	\includegraphics[width=90mm]{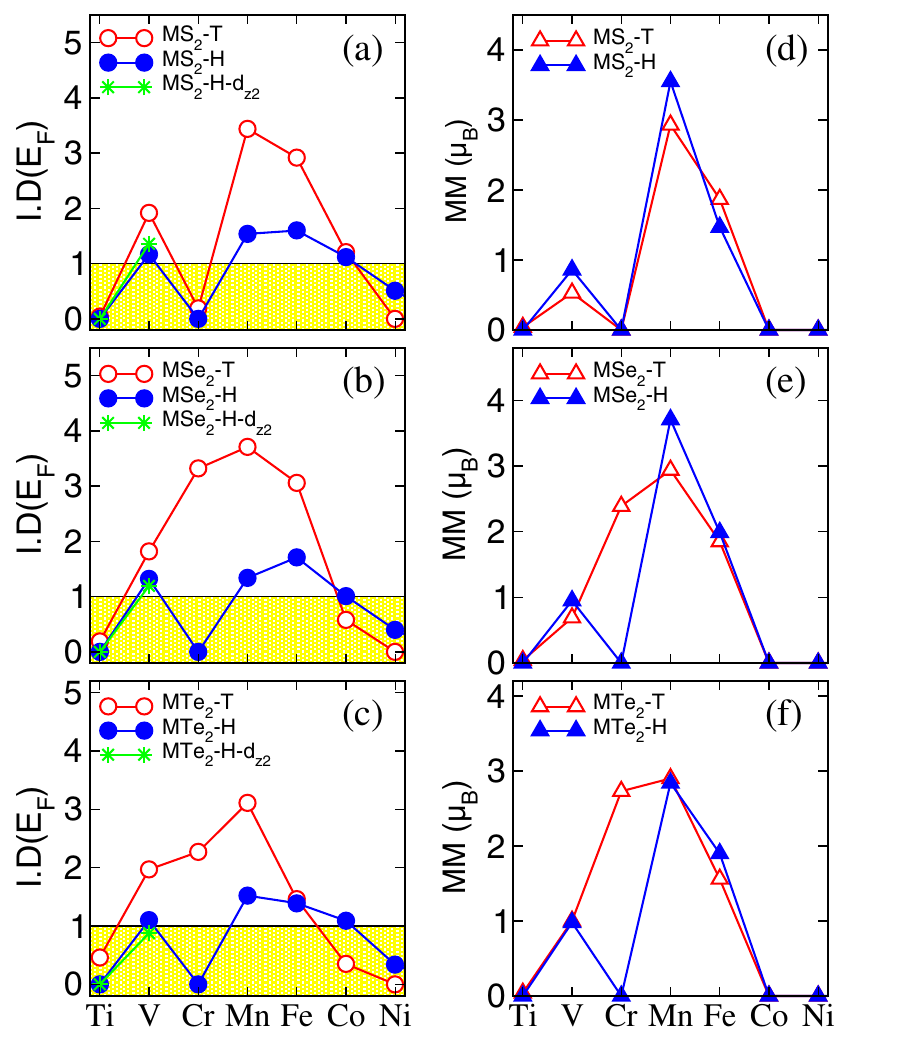}	
	\vspace{-0.1 cm}
	\caption{Stoner criterion for both T and H phase of (a) $M$S$_2$, (b) $M$Se$_2$, and (c) $M$Te$_2$. Calculated
magnetic moments (in units of $\mu_{B}$) of TM atoms for (d) $M$S$_2$, (e) $M$Se$_2$, and (f) $M$Te$_2$. We have also presented the values in the case of $d_{z^{2}}$ correlated subspace with green star points in H-phase.}
	\label{fig:subm6}
\end{figure}

As seen above, most of the metallic TMDC compounds having almost large correlation strength are
expected to display correlation phenomena, such as ferromagnetic ordering.
So, in the following, we discuss the appearance of ferromagnetism
in these materials. Note that ignoring semiconductors T-Ni$X_2$, H-Ti$X_2$, and H-Cr$X_2$, the correlated subspace of
all compounds are partially filled in the non-spin-polarized calculation, thus the Stoner model is well suited to
explain the origin of the ferromagnetism of metallic TMDCs.
Based on this model, the instability
of the paramagnetic state towards ferromagnetic ordering is given by the Stoner criterion
$I.D(E_F) > 1$, where $I$ is the Stoner parameter and $D(E_F)$
is the DOS at the Fermi energy in the nonmagnetic state.
Solving the multiorbital Hubbard model, Stollhoff et al.
found that relationship between the Stoner parameter $I$, Hubbard
$U$, and exchange $J$ is given by $I = 3(U + 6J )/25$ \cite{Stollhoff}.
On the basis of the calculated effective Coulomb parameters  $U$ and $J$
presented in Table ~\ref{table:1} and Table ~\ref{table:2}, the Stoner criterion $I.D(E_F) > 1$ for all compounds
 are presented  in Fig.~\ref{fig:subm6}.  Among the metallic $MX_2$, only H-Ni$X_2$($X$=S, Se, Te), T-Ti$X_2$ ($X$=S, Se, Te), CrS2, and Co$X_2$($X$=Se, Te)
do not satisfy the Stoner criterion. Almost all theoretically predicted ferromagnetic TMDCs fulfill the
Stoner criterion, which is reasonably
consistent with the our results of spin-polarized total
energy calculations and the sizable magnetic moments presented
in right panels of Fig.~\ref{fig:subm6}.
Note that simple Stoner model does not predict correctly the ground state magnetic phase of
H-Co$X_2$ ($X$=S, Se, Te) and T-CoS$_2$. Despite the value of $I.D(E_F)$ for these materials
are larger than 1, they have a non-magnetic ground state.

H-VSe$_2$ and T-MnS$_2$ were among the first TMDCs materials
in which room temperature ferromagnetism were detected experimentally in the monolayer
limit. Despite this, a few works indicated that the ferromagnetism observe in
the TMDCs is not intrinsic and stem from defects or proximity effects\cite{Chua-1,Zhang-2,Vinai}
Also, controversy exists over the Mott insulating behavior in the monolayer of 3$d$ TMDCs.
From the theoretical side, the electronic and magnetic ground state
depend strongly on taking the correct Hubbard $U$ parameter of the $d$
electrons into account in model Hamiltonian or first-principle calculation.
As shown in Fig.~\ref{fig:subm6}, in the case of T-MnS$_2$, the criterion
$I.D(E_F) > 1$ is easily satisfied that explains
why relatively strong ferromagnetic order is observed in the experiment.
Also, it's correlation strength $U/W_b$=2.7 reinforces this idea that the preferred
magnetic order is ferromagnetic.
The situation in the case of H phase of VSe$_2$ is not straightforward.
 We find $I.D(E_F) \sim 1$ which puts system close to the edge of
 paramagnetic to ferromagnetic transition (see left panels of Fig.~\ref{fig:subm6}) .
It may explain why there is no agreement in the
magnetic phase of V$X_2$ materials.

\section{Conclusion}\label{sec4}
We have systematically determined the correlated subspaces and effective on-site and interorbital Coulomb interactions  between localized electrons in 3$d$ $MX_2$ ($M$=Ti, V, Cr, Mn, Fe, Co, Ni and  $X$=S, Se and Te) by employing first \emph{ab initio} calculations in conjunction with a parameter-free cRPA scheme.
These Coulomb interactions not only provide a fundamental understanding
of the correlated phenomena such as magnetic ordering, charge density wave, or Mott phase in TMDCs, but these
effective cRPA parameters can also be used in model Hamiltonians thus increasing the predictive power of model calculations.
 Based on Hubbard $U$, exchange interactions $J$, and electronic structure,
  we find that for most of the TMDC compounds the correlation strength $U/W_b$ are large enough to
 be able to put them in moderate and even strong correlated regime.
Among TMDC materials, the maximum value $U/W_b$=2.7 eV is obtained in MnS$_2$ which is almost three time larger than the corresponding value in elementary Mn. We thus expect electron correlations play an important role in model Hamiltonian studies
of the 3$d$-TMDC.  Since, there is no agreement on the magnetic ordering of these materials, in particular experimentally observed systems like VSe$_2$ and MnS$_2$, we have checked the condition to be fulfilled for the formation of ferromagnetic order by Stoner criterion $I.D(E_F) > 1$.
The results indicate that both Mn$X_2$ ($X$=S, Se) and V$X_2$ ($X$=S, Se) have an intrinsic ferromagnetic behavior in pristine form,
with difference that V-based materials are close vicinity to the critical point separating ferromagnetic from paramagnetic phase
\subsection*{Acknowledgements}
......

\end{document}